\documentstyle[11pt,epsf]{article}
\newcommand{\be}{\begin{equation}}
\newcommand{\ee}{\end{equation}}
\newcommand{\bea}{\begin{eqnarray}}
\newcommand{\eea}{\end{eqnarray}}

\def\lQ{\Lambda_{\rm QCD}}

\def\als{\alpha_{\rm s}}
\def\siml{{\ \lower-1.2pt\vbox{\hbox{\rlap{$<$}\lower6pt\vbox{\hbox{$\sim$}}}}\ }} 
\def\simm{{\ \lower-1.2pt\vbox{\hbox{\rlap{$>$}\lower6pt\vbox{\hbox{$\sim$}}}}\ }}

\newcommand{\Appendix}[1]%
    {%
     \section{#1}%
      }
\def\siml{{\ \lower-1.2pt\vbox{\hbox{\rlap{$<$}\lower6pt\vbox{\hbox{$\sim$}}}}\ }}

\newcommand{\AmS}{{\protect\the\textfont2 A\kern-.1667em\lower.5ex\hbox{M}\kern-.125emS}}

\begin{document}
\begin{center}
\begin{Large}
{\bf A Short Introduction to  Non-Relativistic Effective  Field Theories
\footnote{Invited Talk given at the XXIII International Workshop on the Fundamental Problems of High Energy Physics, Protvino (Russia), June 2000.}} \\[2cm]
\end{Large}
{\large Nora Brambilla}\\ 
{\it Institut f\"ur Theoretische Physik\\ 
Philosophenweg 16, Heidelberg  D-69120, Germany \\
and Dipartimento di Fisica\\
Via Celoria 16, 20133 Milano, Italy}\\[1cm]
\end{center}

\begin{abstract}
I  discuss effective field theories for  heavy bound systems, particularly 
bound systems involving two heavy quarks. The emphasis is on the relevant concepts
and on  interesting physical applications and results. 
\end{abstract}

\section{NON-RELATIVISTIC BOUND SYSTEMS}
In nature there are many  particle bound systems  for which the relative 
velocity $v$ of the 
particle in the  system is small.
Typical examples in the domain of  electromagnetic
 interactions are 
positronium ($e^+ e^-$) and muonium ($e^- \mu^+$), their counterpart 
in the strong interaction domain are  heavy quarkonia ($t\bar{t}$, $b\bar{b}$, 
$b \bar{c}$, $c\bar{c}$) and 
somehow in the middle lie systems like hydrogen, hydrogenoid atoms, 
pionium ($\pi^+ \pi^-$).
All such systems are non-relativistic bound systems. \par
At the beginning of this paper, I will
 use the example of  positronium to show what are the typical (technical)
 problems 
 inherent to a non-relativistic 
bound state calculation, even in a pure perturbative situation 
and I will relate such problems to the existence of several physical scales.
Then, I will discuss the further complications that arise in 
bound state calculations inside 
a strongly coupled theory like QCD.
For the rest of the paper, I will introduce non-relativistic Effective Field Theories (EFT)
 and  I will explain how EFT
greatly simplify the  bound state calculations, 
 both technically and conceptually,
 allowing us to obtain systematically  interesting and new physical results.

\section{AN EXAMPLE of bound state dynamics: POSITRONIUM}
To understand what are the peculiarities of the bound state interaction,
let us first consider positronium.  
Here, the interaction causing the binding 
is the  electromagnetic interaction: QED fully describes this system and the 
coupling constant is the fine structure constant $\alpha$, which is small and 
completely under control in the region of physical interest. 
The positronium energy levels are given  by the  poles 
of the four-point positron-electron  Green function $G=\langle 0 \vert 
 \bar{\psi}_1 {\psi}_2 \bar{\psi}_2 \psi_1\vert 0 \rangle$,
  which in turn is given  as a formal series 
in terms of Feynman  diagrams and thus in terms of $\alpha$.
An integral equation can be written, called Bethe-Salpeter equation, 
whose solution is the four-point Green function $G$ and whose kernel $K$ is the
 subset of two-particle irreducible 
diagrams. Appropriate techniques  allow us to write down formally the 
energy levels as a series of contributions involving
 kernel insertions  on  the zeroth-order Green function 
averaged on the zeroth order wave functions (for a general review see \cite{qedbs}, for 
an explicit calculation see e.g. \cite{ff}).

There is an important difference between the calculation of a 
scattering amplitude  and the calculation of bound state wave functions 
and energy levels. Both are given in terms of Green functions projected 
on initial and final states,
 but in the first case  the initial and final states are on shell, i.e. the wave 
functions describing initial and final particles are free, while in the second 
case the initial and final wave functions are the bound state ones and thus 
bear a dependence on $\alpha$. This last fact produces that 
from the expansion of the energy levels it is not trivial to select
the set of diagrams contributing
 at any given order in $\alpha$. In fact, being the wave function 
$\alpha$-dependent,  the number of vertices in a diagram do not allow 
to trace back the order in $\alpha$ 
of the contribution of the diagram.
Precisely, it happens that: the contribution of each graph is a series in 
$\alpha$; the leading order in $\alpha$ does not follow from the number 
of vertices in the graphs.
For example, diagrams differing only for the number of ladder photons
that they contain, contribute all at  the same leading order in $\alpha$  to 
the energy levels, plus subleading contributions.
Therefore, it is necessary to 
resum an infinite series of diagrams to   complete  an order in $\alpha$ 
in the energy levels. In other words, the bound state calculation are 
'non-perturbative' in the binding interaction.
Moreover, the contribution of a diagram 
depends strongly on the gauge. 
In particular, spurious terms  can be generated and canceled 
only by subsequent  contributions.
In fact, while in a scattering calculation
gauge invariance is manifest term by term at any order of the expansion in 
$\alpha$, in a bound state calculation one has to look for particular subsets
of diagrams and prove that they are gauge invariant. This turns out to be 
quite difficult in practice.\par
The same fact can be explained in the following alternative way.
 Let us consider the calculation 
of the perturbative  corrections to the electron anomalous magnetic moment.
This is a typical scattering amplitude calculation and in the integrals
related to the actual evaluation of the Feynman diagrams  we 
have only one relevant physical scale: the mass $m$ of the electron. On the 
other hand, let us consider the calculation of the 
 positronium energy levels.
This is a typical bound state  calculation and in the integrals
related to the actual evaluation of the Feynman diagrams we 
have three  relevant physical scales: the mass $m$ of the electron,
 the relative momentum $p \sim mv$ and the bound state energy $E \sim 
mv^2$.  For positronium $v \sim \alpha \ll 1 $ and thus 
such scales are quite 
different and get entangled in the calculations causing the Feynman diagrams to 
contribute in a nontrivial way to the perturbative expansion in  $\alpha$, 
as explained above \cite{qedbs}.  \par
This is what happens in the QED calculation of the bound state energy levels.
Alternatively, one can take explicitly advantage of the non-relativistic nature 
of the system and start directly from the non-relativistic reduction of the 
QED bound state problem. In this case the zeroth-order problem is the Schr\"odinger 
equation 
\begin{equation}
\Big (-{\Delta\over m} +V (r) - E\Big ) G^\prime({\bf r}, {\bf r^\prime}, E) =\delta^3({\bf r} 
-{\bf r^\prime}) ,
\label{sch}
\end{equation}
where $V(r)$ is 
the Coulomb potential $V=-{\alpha\over r}$ and $G^\prime$  is now the 
non-relativistic 
Green function. The Coulomb potential term corresponds to  the resummation 
of all the ladder photon contributions. In other words, Eq. (\ref{sch})
provides a resummation of  the most singular terms 
$\sim ({\alpha\over v})^n$
in the Feynman diagrams series. Therefore, this appears to be the  
suitable starting point to study  a non-relativistic bound state like positronium.
However, in order to proceed further and systematically calculate 
perturbative and relativistic corrections,  the non-relativistic 
reduction needs to be under control. A number of problems emerge in any kind 
of naive non-relativistic reduction like:
\begin{itemize}
\item{} {\it Considering retardation (or nonpotential) effects.}
Problems typically arise on trying to go beyond the leading  order 
 (\ref{sch}). Retardation effects, related to low energy photons 
(or gluons in QCD), 
 appear at some point in the reduction. Any precision
perturbative calculation has therefore to deal with them.
However, no systematic procedure  was developed before the effective field theory  approach. 
Such an issue becomes particularly 
relevant in QCD where nonperturbative contributions appear also as nonpotential 
effects (due to very low energy gluons) \cite{voloshin}. 
The lack of  a systematic 
and clear approach to evaluate potential and nonpotential effects,   led 
in the past to many inconsistencies and contradictory statements (about the existence or 
non-existence of the $q\bar{q}$ potential) 
in the frame of the naive non-relativistic   
 reduction.
\item{} {\it Considering  relativistic corrections.}
Problems typically arise on considering the contributions of the relativistic corrections
in the $1/m$ expansion to the energy levels or to the wave function. 
In this case ultraviolet (UV) divergences may arise at some 
order of the   (quantum mechanical) perturbative calculation. The UV divergences 
reflect the fact that the naive non-relativistic limit is just an approximation 
of the field theory, valid for small moment of order $m v$ or smaller.
  
\item{} {\it Calculating contributions in quantum mechanical perturbation 
theory}. As soon as subsequent contributions in the $1/m$ expansion 
are obtained, a question arise about the treatment of such terms.
Have we to include as many terms as possible inside the Schr\"odinger 
equation or have we to treat them perturbatively? 
Such issues are often encountered 
in the literature for example in connection to the spin-spin term
$\sim {\alpha \over m^2} {\bf S_1} \cdot {\bf S_2} \delta^3({\bf r})$. Being such a 
term not bounded from below, the delta function is often substituted by a
suitable regularization and then used  inside the Schr\"odinger
 equation instead that 
in (quantum mechanical) perturbation theory. 
\end{itemize}
In the next Sections I will show how the effective 
field theory formalism provides  a definite solution to these problems
allowing us to fully exploit the simplifications related to the non-relativistic
nature of the interaction. \par
Here, I would like to point out that the difficulties 
discussed above belong  to  bound state calculations in a 
theory fully under control like QED. When we start addressing a 
strongly interacting theory like QCD in the low energy region where 
the heavy quark bound state lie, 
new conceptual complications arise. They are
 related to the existence of nonperturbative 
contributions and to  the presence of a new physical scale, $\Lambda_{\rm QCD}$,
the scale at which nonperturbative contributions become dominant.
In this situation, both the Bethe-Salpeter approach \cite{revbs} and the naive non-relativistic 
reduction \cite{rev} are based  ad-hoc ansatz and 
drastic as well as non systematically improvable  
approximations  and are  biased by ambiguities and indeterminations.
In the case of heavy quark bound states, the effective field theory approach 
provides us with a systematic and under control approach where 
high energy (perturbative) and low energy (nonperturbative) contributions 
are disentangled. This is what I will discuss in the next Sections.\par
More in general we can say {\it that non-relativistic effective field theories 
give an answer to the long-standing problem of how to derive quantum mechanics from 
field theory}.

\section{TAKING ADVANTAGE of the PHYSICAL SCALES}

{\it Let us consider first heavy-light mesons}. In these systems there are 
two relevant physical scales: the mass $m$ of the heavy quark, which is very large 
and in this sense is a 'perturbative' scale, and the scale $\lQ$ which is low and 
is a 'nonperturbative' scale.  Apart from the mass, all the other dynamical 
scales, like the momentum $p$ and the energy $E$, reduce to
$\lQ$ in heavy-light systems. Heavy quark effective theory (HQET) 
\cite{hqet}
is the QCD effective 
field theory that allows us to separate  hard momenta ($p \sim m$)  and soft 
momenta ($p\sim \lQ$). Hard effects can be calculated in perturbation theory; soft effects 
are governed by the spin-flavor symmetry that makes the theory predictive.
Calculations are organized in expansions in $\als(m)$ and ${\lQ/ m}$.    
 The equivalence between HQET and QCD is 
imposed via the 'matching' procedure. Such procedure has been carried out in 
many (equivalent) ways: by imposing the off shell Green functions to be equivalent to 
those of QCD \cite{hq1}, by integrating out the 'antiparticle' degrees of freedom \cite{hq2};
by performing a Foldy-Wouthysen transformation in the QCD Lagrangian \cite{hq3}.
Precisely, the HQET  is organized in a power series in the inverse of the mass of the heavy quark.
Only two kind of terms contribute: 1) terms containing light degrees of freedom (gluons and light 
quarks) only; 2) terms containing a bilinear in the heavy quark field. The size of each term 
is estimated by assigning the scale $\lQ$ to whatever is not a heavy mass in the Lagrangian.
The effective theory thus  provides us with an unambiguous  power counting 
for each operator and allows us to make precise calculations accurate 
up to the desired order of the expansion.\par
{\it The physical situation for bound states formed by TWO heavy quarks is more complicate.}
We have now four scales into  the game: $m$, $p\sim m v$, $E \sim m v^2$ and $\lQ$. 
In QCD it is not true that $v\sim \als$ always.
 Such relation holds true only in the strictly 
{\it perturbative regime defined by the condition: 
$mv \gg mv^2 \gg \lQ$}.  In general $v$ is a nonperturbative quantity, a function of both 
$\als$ and $\lQ$: $v=v(\als,\lQ)$. However, the only relevant fact is that, 
being $v \ll 1$,  the different scales turn out to be  well separated.
The presence of such entangled scales makes perturbative calculations very difficult
as it was explained before. Moreover, it makes even a numerical 
calculation beyond reach. In fact in a lattice simulation 
of a heavy quark bound state, one should have a space-time grid  that is large compared with $1/mv^2$
but with a lattice spacing that is small compared with $1/m$.  To simulate e.g.
$b\bar{b}$ states where $m/mv^2\sim 10$, one needs lattices as large as $100^4$ which
 are beyond our present computing capabilities \cite{nrqcdlat}.

For all these reasons, it is relevant to construct effective field theories 
which are able to disentangle these physical scales.
 One can think about 
constructing an EFT  where $v$ is the small nonperturbative expansion parameter and 
the correction  in the expansion are thus labeled by powers of $v$ (power counting in $v$).
The first effective field theory of this type that has been introduced is
non-relativistic QCD 
(NRQCD) (and NRQED for QED)\cite{nrqcd}.
NRQCD factorizes the scale $m$ but does not deal with the scale hierarchy $mv \gg mv^2$.
Three scales $p, E, \lQ$ 
remain therefore entangled and even in the perturbative regime, complications 
arises in  bound state calculation due to the fact that the power counting becomes
 not univocal beyond leading 
order. The true problem is that NRQCD  does not make explicit the dominance 
of the potential interaction (in the perturbative regime the dominance of the static 
Coulomb force) and the approximate quantum mechanical nature of the system, as 
has been discussed in Sec.2. Then, another EFT that enjoys these characteristics has 
been introduced and is called potential Non-Relativistic QCD (pNRQCD). 
In the next sections I will introduce NRQCD and pNRQCD. For the rest of the paper I will mainly
present and discuss results and predictions of pNRQCD.

\section{NRQCD}
The mass $m$ can be removed from the dynamical scales of the problem using renormalization techniques. 
The idea is to introduce an ultraviolet cutoff $\Lambda$ of the order of the mass $m$ or less
(but much larger than any other physical scale). 
This cutoff  explicitly excludes relativistic heavy quarks from the theory. On the other hand this 
is a sensible choice of cutoff since the physics of heavy quark mesons is dominated by momenta $p \sim mv$.
Since we are dealing with a quantum field theory, the relativistic states we are cutting out 
have a relevant effect on the low-energy physics. But we can compensate for this loss by adding 
new coefficients and new 
local interactions to the Lagrangian. 

To leading order in $1/\Lambda$ such interactions are 
identical to the interactions already present in the theory; beyond leading order we have to include 
nonrenormalizable interactions multiplied by some 'matching' coefficients $c_j$. In 
principle there are infinite such terms to be included, in practice 
we need only few of them. If we aim at an accuracy of order $(p/\Lambda)^n$ we keep in the 
Lagrangian only terms up to and including $O(1/\Lambda)^n$ interactions.
The coupling $m$, $g$, $c_j$ are determined by the requirement that the cutoff 
theory reproduces the results of the full theory up to order $(p/\Lambda)^n$.
In practice is convenient to take $\Lambda \sim m $, 
to  organize the Lagrangian  in powers of $1/m$ making thus explicit 
the non-relativistic  nature of the physical systems and to transform the Dirac field  so as 
to decouple its upper components from its lower components, i.e. separating the 
quark field from the antiquark field (Foldy--Wouthysen transformation).\par
Equivalently, we say that by integrating out, in the sense of the renormalization 
group, high energy degrees of freedom we pass from QCD to NRQCD. The technical 
procedure to achieve this passage is called ``matching'' and the information 
about the integrated degrees of freedom is contained in the 
''matching coefficients''. The matching scale separates the degrees of freedom
 we integrate out in the EFT from those that remain dynamical in the EFT. 
Such scale corresponds to the cutoff of the EFT if the regularization 
procedure is based on cut-off regularization.
At each matching step the non-analytic behaviour in the scale which is integrated 
out becomes explicit in the matching coefficients.
 In this case we are integrating out the mass that becomes an 
explicit parameter in the expansion in powers $1/m^n$ in the Lagrangian  while 
the dependence in $\ln(m/\mu)$ is encoded into the matching coefficients.

Only recently it has been  realized how to perform the matching between QCD and NRQCD in 
dimensional regularization \cite{equiv} and it has been  understood that 
it is the same matching as in HQET (plus four fermion terms)\footnote{Actually only the phenomenological 
use makes the difference between HQET and NRQCD.
In particular, at the level of the Lagrangian, the power counting is
 different. For example, terms with two heavy and two light quark  fields are relevant in 
heavy-light systems but not in heavy-heavy systems and thus are considered in HQET and not in 
NRQCD. Corrispondently, terms with four heavy quark fields are relevant only for heavy-heavy systems 
and not for heavy-light and thus they are considered in NRQCD but not in HQET.
The HQET Lagrangian obtained with matching exploiting procedures different from the Foldy-Wouthysen 
expansion is in fact different from the NRQCD Lagrangian but it is related to it via local
field redefinitions or via the equations of motions. }. 
In fact it is true that, since two scales $p\sim mv$ and $E\sim mv^2$ 
(plus $\lQ$) exist in NRQCD, while only one scale is left  in HQET $p\sim \lQ$, $E\sim \lQ$,
the power counting in the two EFTs is completely different. However, we  should 
not  be mistaken by this fact.
The key  point here is that in the matching,
the power counting of the effective theory plays no role at all, the 
only relevant thing  being that all the other dynamical scales 
(present in the effective theory) are much lower than the matching scale.
The matching computation in the 
effective theory in dimensional regularization  is zero, which greatly simplifies the calculation.
Dimensional regularization,
and the $\overline{{\rm MS}}$
 scheme is a useful tool that we will always imply  in 
in the following.\par
Since the 
scale of the mass of the heavy quark is perturbative, the scale $\mu$
of the matching from QCD to NRQCD, $mv <\mu < m$, lies also in the perturbative regime. Then, it is possible  ``to integrate out''
 the  hard scale  by comparing on shell amplitudes,
expanded order by order in ${1/m}$ and in $\alpha_s$,  in QCD and in NRQCD.  
The difference is  encoded into  the matching coefficients that 
typically depend non-analytically on the scale 
which has been integrated out (like $\ln(\mu/m)$),  in this case $m$. 
We call this type of matching {\it perturbative matching} because it is calculated via an
expansion  in $\als$.

Up to order ${1/m^2}$ the Lagrangian of NRQCD \cite{nrqcd,nrqcdrev,match} reads:
\begin{eqnarray}
& &\ L_{\rm NRQCD} = \psi^\dagger\left(iD_0 +  c_2  {{\bf D}^2\over 2  m} + 
 c_4  {{\bf D}^4\over 8  m^3} 
+  c_F  g { {\bf  S}\cdot {\bf B} \over  m } +  c_D  g { {\bf D}\cdot{\bf E} 
 - {\bf E}\cdot{\bf D} \over 8  m^2}\right. \label{NRQCD}\\ 
& & \left. + i  c_S  g {{\bf S}\cdot({\bf D}\times{\bf E} - 
{\bf E} \times {\bf D}) \over 4  m^2}\right) \psi
+ \hbox{\,  antiquark terms} + \hbox{\, terms with light quarks} \nonumber\\
& & 
-{ b_1 \over 4} F_{\mu\nu}^a F^{a\,\mu\nu}  
+ {b_2\over  m^2} F_{\mu\nu}^a D^2 F^{a\,\mu\nu}
 + { b_3 \over m^2} g  f_{abc} F_{\mu\nu}^a  F_{\mu\alpha}^b F_{\nu\alpha}^c \nonumber \\
& & + { d_1 \over  m^2} \psi^\dagger \psi \chi^\dagger \chi   
+ { d_2  \over  m^2}  \psi^\dagger {\bf S} \psi \chi^\dagger {\bf S} \chi 
+ { d_3 \over  m^2} \psi^\dagger  T^a  \psi \chi^\dagger  T^a  \chi 
+ \!{ d_4 \over  m^2} 
\psi^\dagger  T^a {\bf S}  \psi \chi^\dagger  T^a {\bf  S}  \chi, \nonumber
\end{eqnarray}
$\psi$ and $\chi$ being respectively the quark and the antiquark field; $D^\mu$
is the covariant derivative, ${\bf E}$ and ${\bf B}$ are chromoelectric and chromomagnetic 
fields, $F^{\mu\nu}$ is the gluon field strength, ${\bf S}$ is the total spin,
$T^a$ is the color generator, $g$ is the coupling constant, $\als={g^2/ 4 \pi}$.  
$c_i,b_i,d_i$ are matching coefficients. 
The gluonic part  in the third line comes from the (heavy quark) vacuum polarization 
while the last line contains the four quark operators.\par  
Since the mass $m$ is explicitly removed from the dynamics, a lattice evaluation 
of a typical heavy quark bound system like bottomonium may now take place at 
lattice spacings larger by a factor $1/v$. This reduces  the needed size of the 
lattice by a factor $1/v^4$, which is approximately  a factor 100 for the 
$1S$ of $b \bar{b}$. 

The matching coefficients  depend on $\mu$ and $m$ and 
are known in the literature at a different level of 
precision \cite{match,amoros,nrqcdmatch}. The $\mu$ dependence in the matching 
coefficients cancels against  the $\mu$ dependence of the operators in the 
Lagrangian.
\begin{figure}
\makebox[1cm]{\phantom b}
\put(75,0){\epsfxsize=8.5truecm \epsfbox{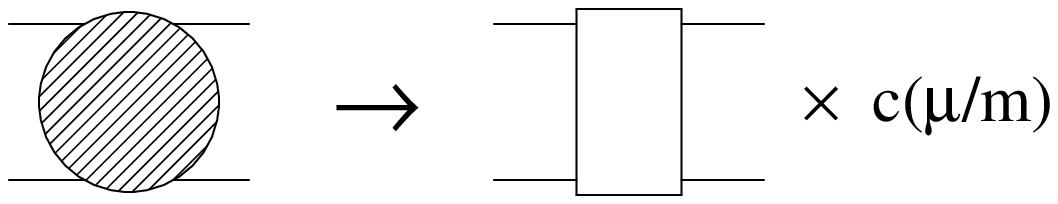}}
\put(85,-15){QCD}\put(175,-15){NRQCD}
\caption{\it Matching procedure: on shell green function expanded in $\als$ and $1/m$ are equated 
in QCD and in NRQCD. The difference is encoded in the matching coefficients.}
\end{figure}
In particular,
 $ b_1$  fixes the new $\beta$-function \cite{match}. 
Let us see in which way. Suppose that in QCD we have $n_f$ 
flavors. Then in NRQCD we have $n_f^\prime=n_f-1$ relativistic flavors. 
Consequently the 
$\overline{{\rm MS}}$
 running coupling constant
 in NRQCD is expected to run according to $n_f-1$ flavors.
In fact in (\ref{NRQCD}) the gluonic kinetic term does not have the standard normalization but 
it is multiplied by $b_1$. This can be recovered by a field redefinition of the gluon field.
Since the remaining gluon  field in $L_{\rm NRQCD}$ are multiplied by $g$, this is equivalent to 
make in these terms the change
\begin{equation}
\als \rightarrow \als \Big (1-{\als\over 3 \pi} T_F \ln{m^2\over \mu^2}\Big )^{-1}= {\als\over b_1}.
\label{change}
\end{equation}
At one loop this corresponds to substituting the running coupling constant of $n_f$ flavors 
with the running coupling constant of $n_f-1$ flavors.
Therefore, the $\als$ multiplying the gluon fields is understood at the scale $\mu$ running according 
to $n_f-1$ flavors.
In all the matching coefficients not multiplying gluon fields, 
 $\als^{n_f}$ can be substituted with $\als^{n_f-1}$ (being the difference in higher orders in $\als$)
and the scale turns out to be fixed to $m$: $\als=\als(m)$.\par

The coefficients $ b_2, \dots  \dots$ and $d_1 \dots$  are known at 1-loop, 
\cite{equiv,match} and $c_F$ at two loops (in the anomalous dimension)
\cite{amoros},
\begin{equation}
 c_F = \left( { \als(m)\over  \als(\mu) } \right)^{\gamma_0 /2\beta_0} 
\left[ 1 + {\als(m)\over 2\pi} (C_A+C_F)    + {\als (m) - \als(\mu) \over 4\pi} 
{\gamma_1\beta_0 - \gamma_0\beta_1 \over 2 \beta_0^2} \right]
\end{equation}
being  $ \gamma_0$, $ \gamma_1$  the anomalous dimensions,
and \cite{Bauerm}
\begin{equation}
c_D  = 
\left({20\over 13} + {32\over 13} {C_F\over C_A} \right) 
\left[1 - \left( {\als(m)\over \als(\mu) } \right)^{13 C_A / 6\beta_0} \right]
+ \left( {\als(m)\over \als(\mu) } \right)^{2 C_A / \beta_0}, 
\end{equation}
where we have presented the renormalization group (RG) improved results.
Notice that the one loop correction (finite part) 
to the chromomagnetic coefficient $c_F$  turns out to 
give a large contribution,
 of order $15 \%$ in $b\bar{b}$ states and of order $30\%$ in $c \bar{c}$ states.
For what concerns the evaluation of the spin splittings therefore  the matching coefficient 
is more important than the subsequent $O(v^4)$ corrections.

The effective field theory must have the symmetries of the fundamental 
theory it is derived from. Hence NRQCD/HQET must have the symmetries of QCD,
in particular Lorentz invariance. Such symmetry is implemented in a nontrivial
way as a reparametrization invariance \cite{rep}. 
In practice, thus Lorentz invariance is
 realized through  the existence of relations between the matching coefficients,
precisely: $ c_2 = c_4 = 1$, $ c_S = 2c_F -1$. 
We stress that :
\begin{itemize}
\item{} The matching  from QCD to NRQCD is perturbative.
\item {} In NRQCD two dynamical scales are still present, $mv$ and 
$mv^2$. Because of the existence of these two scales, the size of each term in 
(\ref{NRQCD}) is not unique. Counting rules have been given\cite{nrqcdlat2} to estimate 
the leading size of 
the matrix elements of the operators in (\ref{NRQCD}) 
but still they do not have a unique power counting  but they also contribute to 
subleading orders in $v$. No explicit  power counting rules  to systematically 
incorporate  subleading effects exist, even in the  perturbative situation.
The entangling of the two surviving dynamical scales makes still very difficult
the evaluation of the Feynman diagrams in any perturbative calculation. 
\item{} It is a consequence of the previous observation that the 
 NRQCD Lagrangian contain, mixed,  both potential and retardation corrections. Indeed both 
potential (or soft (S)) and ultrasoft (US) degrees of freedom remain dynamical in NRQCD, 
while the scale of the mass (hard) has been integrated out. 
\item{} The rest mass $m$ has been removed from the quark energies, allowing for much coarser lattice 
than in the Dirac case. The quark and the antiquarks have been decoupled and thus the quarks 
Green function satisfies a Schr\"odinger like equation
\begin{equation}
\Big (i D_0 + {{\bf D}^2 \over 2m} + \dots \Big ) G^{\rm NRQCD} (x,x^\prime) = \delta^4(x-x^\prime)
\label{schnrqcd}
\end{equation}
that is easily solved numerically as an initial value problem \cite{nrqcdlat}.\par
However, in the most part  of  the NRQCD lattice evaluations 
only mean field contributions (devised to correct for the large tadpole
terms in lattice perturbation theory\cite{nrqcdlat2}) are 
considered while for the rest the matching coefficients  are taken  at 
the tree  level. Such a procedure appears to be inconsistent: once the power counting in $v$ is 
established order $O(\als)$ corrections to the matching coefficients
have to be included via at least one loop calculations. Notice also 
that, as discussed above, the loop corrections to the matching coefficients may be  large 
\cite{nrqcdmatch}.
\end{itemize}
In NRQCD  still the dominant role of the potential as well as the quantum 
mechanical  nature of the problem (Schr\"odinger equation) 
are not  maximally exploited: 
the NRQCD quark Green function satisfies in fact an equation (\ref{schnrqcd}) in which potential
and retardation contributions appear still mixed together \cite{acebaf,others}.\par  
This goal  is achieved  by another  EFT called potential Non-Relativistic QCD (pNRQCD) 
\cite{prepnrqcd,pnrqcd,pnrqcdrev,potpnrqcd}.

\section{pNRQCD}

We want to build an effective field theory that describes the low energy region of the non-relativistic 
bound state, i.e. we want an  EFT where only the ultrasoft degrees of freedom 
remain dynamical while all the other nonrelevant scales have been integrated out.
This fixes two ultraviolet cutoffs for the new EFT: $\Lambda_1$ and $\Lambda_2$. 
The first one satisfies $mv^2 \ll \Lambda_1 \ll mv$ and is the cutoff of the energy of the 
quarks and of the energy and the momentum of the gluons, the second one satisfies 
$mv \ll \Lambda_2 \ll mv^2 $ and is the cutoff of the relative 
three-momentum ${\bf p}$  of the heavy 
quark bound system. This is the maximal reduction (=maximal number of degrees of freedom that 
we can make nondynamical) we can achieve in the description of the 
heavy quark bound state:
in fact the heavy quarks with US energy have a soft three-momentum 
due to their non-relativistic dispersion relation.

First we want to consider a situation where {\it the matching to the new EFT is still perturbative},
i.e. a situation in which the soft scale is still a perturbative scale: $mv \gg \lQ$.
The nonperturbative matching situation: $mv \sim \lQ$, where the matching cannot be carried 
out via an expansion in $\als$ is discussed  in Sec. 12. 
Roughly speaking, we can say that the 
lowest excitations of quarkonium belong to the first situation while the excited states 
belong to the second situation. Indeed, the typical radius of the bound state is proportional
to the inverse of the soft scale $r \simeq 1/mv$ and thus for the lowest states 
the condition $mv \ll \lQ$ may be fulfilled \cite{pnrqcdrev}.
\begin{figure}
\makebox[1cm]{\phantom b}
\put(75,0){\epsfxsize=8.5truecm \epsfbox{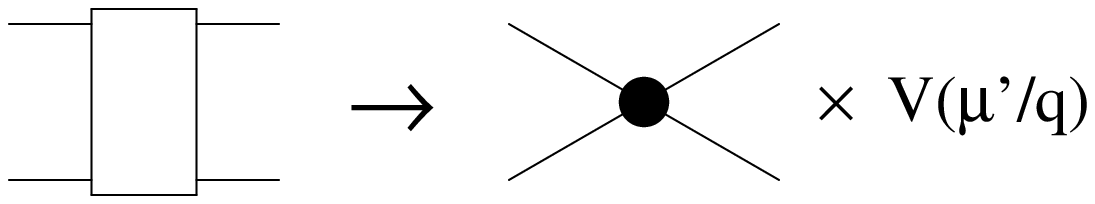}}
\put(85,-15){NRQCD}\put(165,-15){pNRQCD}
\caption{\it Matching procedure: off shell Green functions expanded in $\als$, $1/m$ and 
in the multipole expansion are equated 
in NRQCD and in pNRQCD. 
The difference is encoded in the potential matching coefficients that 
depend nonanalytically on the scale that has been integrated out, in this case $r$.}
\end{figure}

\section{pNRQCD for $mv \gg \lQ$}
We denote by ${\bf R}\equiv {({\bf x}_1+{\bf x}_2})/2$ the center of mass of the $Q\bar{Q}$ 
system and 
by  ${\bf r\equiv {\bf x}_1 -{\bf x}_2}$ the relative distance. 
 At the scale of the matching $\mu^\prime$ 
($mv \gg \mu^\prime \gg mv^2, \lQ$) we have still quarks and gluons.
The effective degrees of freedom are: $Q\bar{Q}$ states (that can be decomposed into 
a singlet $S({\bf R},{\bf r},t)$ and an octet $O({\bf R},{\bf r},t)$
under color transformations) with energy of order of the next relevant 
scale, $O(\Lambda_{QCD},mv^2)$ and momentum   ${\bf p}$ of order $O(mv)$,  plus 
ultrasoft gluons $A_\mu({\bf R},t)$ with energy 
and momentum of order  $O(\lQ,mv^2)$. Notice that all the  gluon fields are multipole 
expanded. The Lagrangian is then  an expansion 
in the small quantities  $ {p/m}$, ${ 1/r  m}$ and in   
$O(\Lambda_{\rm QCD}, m v^2)\times r$.

\subsection{The pNRQCD Lagrangian}
At the next-to-leading order (NLO) in the multipole expansion,
 i.e. at $O(r)$, we get \cite{pnrqcd,potpnrqcd} 
\begin{eqnarray}
& & 
 L^{(1)}_{\rm pNRQCD} 
=   -{1\over 4}  F_{\mu\nu}^a F^{\mu\nu\,a} 
+{\rm Tr} \left\{{\rm S}^\dagger \left( i\partial_0 - {{\bf p}^2\over  m } 
- V_s
-\sum_{n=1} { V^{(n)}_s\over  m^n} \right){\rm S} \right\} \label{pnrqcd}\\
& & + {\rm Tr} \left\{ {\rm  O^\dagger} \left( i{ D_0} - {{\bf p}^2\over  m}-V_o 
- \sum_{n=1} { V^{(n)}_o\over  m^n} \right){\rm  O} \right\}\nonumber \\
& & 
+ {g}  V_A{\rm Tr} \left\{  {\rm  O^\dagger} {\bf r}\cdot{ \bf E}\,{\rm  S} 
+ {\rm  S^\dagger} {\bf r}\cdot{ \bf E} \,{\rm  O} \right\} 
+ { g} { V_B\over 2} {\rm Tr} \left\{  {\rm  O^\dagger} 
{ \bf r}\cdot{ \bf E}\,{\rm  O}
+ {\rm  O^\dagger} {\rm  O }{ \bf r}\cdot{ \bf E} \right\}.\nonumber
\end{eqnarray} 
All the gauge fields in Eq. (\ref{pnrqcd}) are evaluated 
in ${\bf R}$ and $t$. In particular ${\bf E} \equiv {\bf E}({\bf R},t)$ and 
$iD_0 {\rm O} \equiv i \partial_0 {\rm O} - g [A_0({\bf R},t),{\rm O}]$. 
The quantities denoted by $V_j$ are the matching coefficients. 
They are functions of $\mu, \mu^\prime, {\bf r}, {\bf p}, {\bf S}_1, {\bf S_2}$.
We call $V_s$ and $V_o$ the singlet and octet static matching potentials respectively.
The equivalence of pNRQCD to NRQCD, and hence to QCD, is enforced
by requiring the Green functions of both effective theories to be equal (matching).
In practice, appropriate off shell amplitudes are compared in NRQCD and in pNRQCD,
order by order in the expansion in $1/m$, $\alpha_s$  and in the multipole expansion.
The difference is encoded in  potential-like matching coefficients 
that depend non-analytically on the scale that has been integrated out (in this case ${\bf r})$.\par
\begin{figure}[htb]
\makebox[0.5cm]{\phantom b}
\epsfxsize=1.5truecm \epsfbox{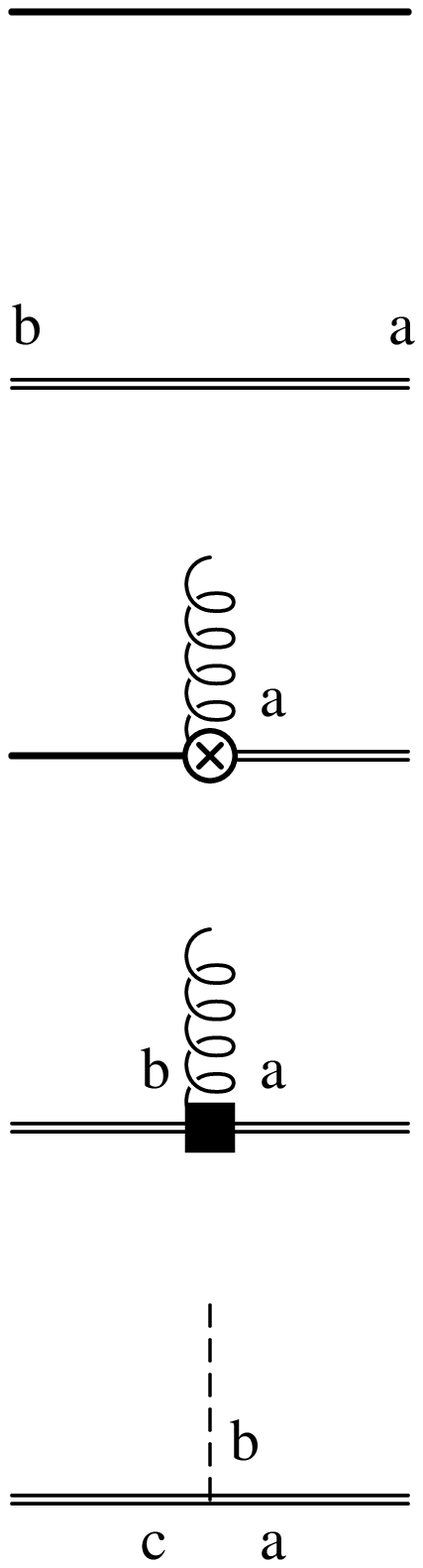}
\put(50,160){\small $= \theta(T) e^{\displaystyle -i V_s T}$}
\put(210,160){\small singlet propagator}
\put(50,126){\small $= \theta(T)\Bigg(e^{\displaystyle -i V_o T} 
e^{\displaystyle - i g \int_{-T/2}^{T/2} \!\!\! dt 
                 \, A_0^{\rm adj}}\Bigg)_{ab}$} 
\put(210,126){\small octet propagator}
\put(50,86){\small $= i g V_A \displaystyle\sqrt{T_F\over N_c} {\bf r} \cdot {\bf E}^a$}
\put(245,86){\small singlet--octet vertex}
\put(50,46){\small$= i g \displaystyle{V_B\over 2} d^{abc} {\bf r} \cdot {\bf E}^c$}
\put(245,46){\small octet--octet vertex}
\put(50,6){\small $=  g f^{abc}$}\put(210,6){\small Coulomb octet--octet vertex}
\caption{ \it Propagators and vertices of the pNRQCD Lagrangian (at order $1/m^0$ and 
at the NLO in the multipole expansion).}
\label{due}
\end{figure}

At the leading order (LO) in the multipole expansion, 
the equations of motion of the singlet field is the 
 Schr\"odinger equation 
\begin{equation}
i\partial_0 S= \left ({{\bf p}^2\over m} + V_s(r) \right ) S.
\label{schp}
\end{equation}
Therefore pNRQCD has made explicit the dominant role of the potential 
and the quantum mechanical nature 
of the bound state. In particular both the kinetic energy and the potential 
count as $mv^2$ in the $v$ power counting.
The leading order\footnote{The non-relativistic limit described by the 
Schr\"odinger equation with the static potential is called 'leading order' (LO); contributions
corresponding  to corrections of order $v^n$  to this limit are called ${\rm N}^n{\rm LO}$. LL
means 'leading log'.} 
  problem in fact reduces to the usual Schr\"odinger equation;
the actual bound state  calculation turns out to be very similar to a standard 
quantum mechanical calculation, the only difference being that the wave 
function field couples to US gluons in a field theoretical fashion. 
 From the solution of the Schr\"odinger equation come the leading order propagators for the singlet and the 
octet state, while the vertices come from the interaction terms at the NLO 
in the multipole expansion. In  Fig.(\ref{due}) you find such 
the Feynman rules in the static case
(useful for the matching). For calculations inside pNRQCD the kinetic term ${\bf p}^2/m$ has to 
be included in the singlet and octet propagators.\par  
The last line of (\ref{pnrqcd})
contains  retardation (or nonpotential) effects that 
start at the NLO in the multipole expansion. At this order the nonpotential
effects come from the singlet-octet interaction and the octet-octet interaction 
mediated by a ultrasoft chromoelectric 
field.\par
 Recalling that ${\bf r} \simeq 1/mv$ and that the operators count like the next relevant 
scale, $O(mv^2,\lQ)$, to the power of the dimension, it follows that  
each term in  the pNRQCD Lagrangian has a definite power counting.  This feature makes 
$L_{\rm pNRQCD}$ the most suitable tool for a bound state calculation: being interested 
in knowing the energy levels up to some power $v^n$, we just need to evaluate the contributions
of this size in the Lagrangian. 

We stress that
\begin{itemize}
\item{pNRQCD  is equivalent to QCD.}
\item{ pNRQCD  has explicit potential terms and thus   it embraces a description of heavy quarkonium 
in terms of potentials.}
\item{pNRQCD  has explicit dynamical ultrasoft gluons  and thus  it describes nonpotential 
(retardation) effects.
 US gluons are incorporated in a second-quantized, gauge-invariant and systematic way.
In the appropriate dynamical situation $\lQ\ll mv^2$ and $\lQ \sim mv^2$
 these give nonpotential terms of nonperturbative 
nature, like the term due to the gluon condensate \cite{voloshin}.
From the power counting it follows that the interaction of quarks with ultrasoft gluons 
is suppressed in the lagrangian by $v$ with respect to the LO (by $gv$ if $mv^2 \gg \lQ$).}
\item{ The power counting is unambiguous. Being all the scales disentangled, the perturbative 
calculation of the bound state is considerably simplified. }
\item{ Calculations can be performed  systematically in the $v$ expansion
and can be improved at the desired order.}
\item{Perturbative (high energy) and nonperturbative (low energy) contributions are disentangled.
we can therefore systematically parameterize the nonperturbative contributions that we are not able 
to evaluate directly.}
\item{Lorentz invariance is again implemented through reparameterization 
invariance. In this case this implies the existence of relations between 
the potential matching coefficients \cite{relpot}.}
\item{We have a definite prescription (power counting) on the calculations of 
the contribution of the various relativistic 
corrections to the energy levels. 
We have to  solve first the Schr\"odinger equation with the static potential. All 
the rest, if it is suppressed by the power counting,
 has to be computed using quantum mechanical perturbation theory on the leading order Schr\"odinger 
solution. E.g. the spin-spin term proportional to the delta function I considered in Sec. 2 
{\it has to be calculated} in quantum mechanical perturbation theory and {\it has not} to be 
included in the Schr\"odinger equation.}
\end{itemize}

In pNRQCD the potentials are defined upon integration of all the scales {\it up to the ultrasoft 
scale $mv^2$} and 
are understood in the modern acception of matching coefficients:  
they are dependent on the scale of the matching $\mu^\prime$.  Of course, such scale dependence is
canceled in the energy levels by the contribution of the ultrasoft gluons which are cutoff at 
the same scale $\mu^\prime$.  In particular pNRQCD provides us with a well defined procedure 
to obtaining the matching 
potentials  via the matching at any order of the perturbative expansion. This is particularly 
relevant in QCD, where the  calculation of the static potential is unclear in a naive 
perturbative frame, see e.g.\cite{pot,potpnrqcd}. 
Up to now the EFT approach is the only one able 
to supply a perturbative definition of the static potential at all orders, in particular 
to enable a concrete calculation at three loops (LL)
and consequently  a perturbative calculation of the energy levels at NNNLL. This is an 
example where the EFT seems {\it not only useful but necessary}.

\subsection{The $Q\bar{Q}$ potential in perturbation theory}

The existence of the different physical scales makes even a purely perturbative definition of the 
static $Q\bar{Q}$ potential not free from complications. Let us consider the energy of the static 
quark sources
$ E_s(r) =  \displaystyle\lim_{T\to\infty} {i \over T} \ln  \langle W_\Box \rangle$, 
(being $W_\Box$  the static Wilson loop of size ${\bf r} \times T$,  and 
the symbol $\langle ~~ \rangle$ being the average over the gauge fields), which is usually considered 
as a definition of the static potential.
 At three loops $E_s$  shows infrared divergences\cite{pot,potpnrqcd}. 
These singularities may indeed be regulated,  upon resummation of a certain 
class of diagrams, which give rise to a sort of dynamical cut-off provided by 
the difference between the singlet and the octet potential. However, such a dynamical 
scale is of the same order of the kinetic energy 
for quarks of large but finite mass and, therefore, should not 
be included into a proper definition of 
the static potential in the sense of the Schr\"odinger equation.
This is similar 
to what happens for the Lamb shift in 
QED at order $1/m^2$. In QCD this effect  calls, even in the definition of the 
static potential,  for a rigorous treatment of the bound-state scales. 
To address the multiscale dynamics of the heavy quark bound state, 
the concept of effective field theory  turns out to be  not only helpful  but actually
necessary.\par
We answer these questions  by  performing  explicitly the  singlet
matching  at order $1/m^0$ and at  the NLO in the multipole expansion.  \par
The matching can be done once the interpolating fields for $S$ and $O^a$ have been identified in NRQCD. 
The former need to have the same quantum numbers and the same transformation properties as the latter. 
The correspondence is not one-to-one. Given an interpolating field in NRQCD, 
there is an infinite number of combinations of singlet and octet wave-functions 
with ultrasoft fields, which have the same quantum numbers 
and, therefore, have a non vanishing overlap with the NRQCD operator. However, the operators in pNRQCD 
can be organized according to the counting of the multipole expansion. 
For instance, for the singlet we have  
\begin{equation}
\chi^\dagger({\bf x}_2,t) \phi({\bf x}_2,{\bf x}_1;t) \psi({\bf x}_1,t) =  Z^{1/2}_s(r) S({\bf R},{\bf r},t) 
+ Z^{1/2}_{E,s}(r) r \, {\bf r}\cdot{\bf E}^a({\bf R},t) O^a({\bf R},{\bf r},t) + \dots,  
\label{Sdef}
\end{equation}
and  for the octet 
\begin{eqnarray}
\!\!\!\chi^\dagger({\bf x}_2,t) \phi({\bf x}_2,{\bf R};t) T^a \phi({\bf R},{\bf x}_1;t) \psi({\bf x}_1,t) 
&=& Z^{1/2}_o(r) O^a({\bf R},{\bf r},t) \label{Odef}\\
&+& Z^{1/2}_{E,o}(r) r \, {\bf r}\cdot{\bf E}^a({\bf R},t) S({\bf R},{\bf r},t) + \dots, 
\nonumber
\end{eqnarray}
$
\phi({\bf y},{\bf x};t)\equiv {\rm P} \exp \{ ig 
\int_0^1 \!\! ds \, ({\bf y} - {\bf x}) \cdot {\bf A}({\bf x} - s({\bf x} - {\bf y}),t) \}$, $Z_i$ being 
normalization factors.
These operators guarantee a
leading overlap with the singlet and the octet wave-functions respectively. Higher order corrections are 
suppressed in the multipole expansion. 
The expressions for the octet can be made manifestly 
gauge-invariant by inserting a chromomagnetic field in place 
of the pure color matrix. The fact that the matching can be done in a completely gauge-invariant 
way enables us to generalize pNRQCD to the case in which $\lQ \simeq mv$, i.e. to the nonperturbative
 matching, cf. Sec. 12.\par 
Now, in order to get the singlet potential, we compare 4-quark Green functions. From (\ref{Sdef}),
we take in NRQCD\begin{equation}
I = \delta^3({\bf x}_1 - {\bf y}_1) \delta^3({\bf x}_2 - {\bf y}_2) \langle W_\Box \rangle , 
\label{vsnrqcd}
\end{equation}
 and we equate  
(\ref{vsnrqcd}) to the singlet propagator in pNRQCD
at NLO in the multipole expansion (cf. Fig. 4 for a diagrammatic representation)
\begin{eqnarray}
& &
\!\!\!\!\!\!\!\!\!
I = Z_s(r) \delta^3({\bf x}_1 - {\bf y}_1) \delta^3({\bf x}_2 - {\bf y}_2) e^{-iTV_s(r)} \times
\label{vspnrqcdus}\\
& &
\!\!\!\!\!\!\!\!\!\!\!\!\Bigg( 1 -{ g^2 \over N_c} T_F V_A^2 (r)\int_{-T/2}^{T/2} \!\!\! dt 
\int_{-T/2}^{t} \!\!\! dt^\prime e^{-i(t-t^\prime)(V_o-V_s)} 
\langle {\bf r}\cdot {\bf E}^a(t) \phi(t,t^\prime)^{\rm adj}_{ab}
{\bf r}\cdot {\bf E}^b(t^\prime)\rangle \Bigg),
\nonumber
\end{eqnarray}
where $\phi^{\rm adj}$ is a Schwinger (straight-line) string in the adjoint representation 
and fields with only temporal argument are evaluated in the centre-of-mass coordinate.
\begin{figure}[htb]
\makebox[4cm]{\phantom b}
\epsfxsize=7.7truecm \epsfbox{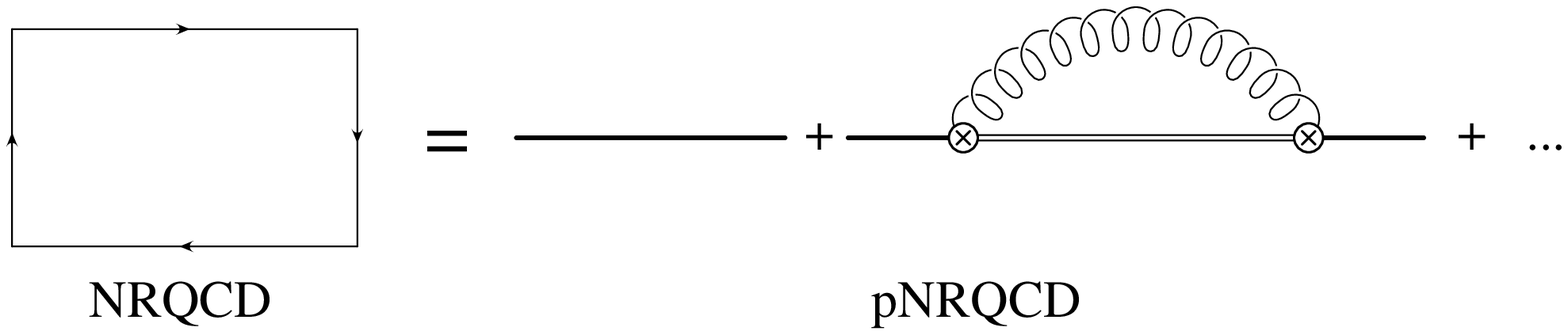}
\vskip -0.4truecm
\caption{\it The matching between the Wilson loop in NRQCD and the singlet propagator in pNRQCD.}
\vskip -0.05truecm
\end{figure}

Comparing Eqs. (\ref{vsnrqcd}) and (\ref{vspnrqcdus}), one gets at the 
next-to-leading order in the multipole expansion the singlet wave-function normalization $Z_s$ 
and the singlet static potential $V_s$. $V_A$ and $V_o$ must have 
been previously obtained from the matching of suitable operators, but for the present purposes 
we only need the tree-level values: $V_A = 1$ and $V_o = (C_A/2-C_F)\alpha_{\rm s}/r$. 
Let us concentrate here on the matching potential $V_s$. 
By substituting the chromoelectric field correlator in Eq. (\ref{vspnrqcdus}) 
with its perturbative expression we obtain  at the next-to-leading order in the 
multipole expansion and at order $\alpha_{\rm s}^4 \ln \alpha_{\rm s}$
\begin{equation}
V_s(r) = E_s(r)\big\vert_{\rm 2-loop+NNNLL} 
+ C_F {\alpha_{\rm s}\over r} {\alpha^3_{\rm s}\over \pi}
{C_A^3\over 12} \ln {C_A \alpha_{\rm s} \over 2 r \mu}.  
\label{vsu0}
\end{equation} 
The 2-loop contribution to $E_s$ has been calculated in \cite{twoloop}. The NNNLL contributions 
arise from the diagrams studied first in \cite{pot} and shown below.
\begin{figure}[h]
\makebox[0cm]{\phantom b}
\put(40,0){\epsfxsize=4.5truecm \epsfbox{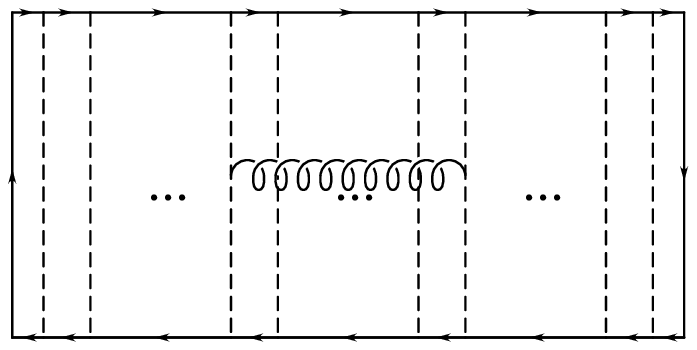}}
\put(185,25){$= -\displaystyle{C_FC_A^3 \alpha_{\rm s}^4 \over 12 \pi r} 
\ln {C_A \alpha_{\rm s}\over 2} + \displaystyle O\left( 1/T \right)$ }
\put(130,5){}
\end{figure}
Note that $V_s$ and $E_s$ would coincide 
in QED and that, therefore, this difference here is a genuine QCD feature.
Such difference is switched on  at NLO in the multipole expansion. 
An explicit calculation gives \cite{potpnrqcd}
\begin{eqnarray}
& &  
\!  V_s(r) \equiv  - C_F {\alpha_{V}(r,\mu^\prime) \over r}, 
\!\label{newpot}\\
& & 
\! {\alpha}_{V}(r, \mu)=\alpha_{\rm s}(r)
\left\{1+\left(a_1+ 2 {\gamma_E \beta_0}\right) {\alpha_{\rm s}(r) \over 4\pi}\right.  
+{\alpha_{\rm s}^2(r) \over 16\,\pi^2}
\bigg[\gamma_E\left(4 a_1\beta_0+ 2{\beta_1}\right)+\left( {\pi^2 \over 3}+4 \gamma_E^2\right) 
{\beta_0^2} 
+ a_2\bigg] \left. \right. \nonumber \\
& & 
 \left. \quad  + {C_A^3 \over 12}{\alpha_{\rm s}^3(r) \over \pi} \ln{ r \mu^\prime}\right\},
\nonumber
\end{eqnarray}
where $\beta_n$ are the coefficients of the beta function ($\alpha_{\rm s}$ is in the $\overline{\rm MS}$ scheme), 
and $a_1$ and $a_2$ were given in \cite{twoloop}. 
We see that the interpretation of the potentials as matching coefficients in pNRQCD 
implies that the Coulomb potential is not simply coincident with the static energy $E_s$.
The Coulomb potential turns out to be sensitive to the ultrasoft scale. The same 
happens with the other potentials (like $V_o$ or the potentials that bear corrections in $1/m^n$)
 that can equally be calculated via the matching procedure.\par

\subsection{The singlet potential in the situation $ mv \gg \Lambda_{\rm QCD} \gg mv^2$}
Since in this situation there is a physical scale ($\lQ$)
above the US scale, a potential can be properly defined only once this scale has been integrated out.
At the NLO in the multipole expansion we get 
\begin{eqnarray}
& &
\!\!\!\!\!\!\!\!\!\!\!\!\!\!\!V_s(r) = -C_F {\alpha_{V}(r,\mu^\prime)\over  r }
 -i{g^2 \over N_c}T_F V_A^2(r){r^2\over 3} \int_0^\infty\!\!\!\! dt 
e^{-it(V_o-V_s)} 
\langle {{\bf E}^a}(t)\phi(t,0)^{\rm adj}_{ab}{\bf E}^{b}(0) \rangle(\mu^\prime).
\nonumber \\
& & \label{vsnp}
\end{eqnarray}
Therefore, the heavy quarkonium static potential $V_s$ is given in this situation 
by the sum of the purely 
perturbative piece calculated in Eq. (\ref{newpot}) and a  term carrying   
 nonperturbative contributions (contained into non-local gluon field correlators). 
This last one can be organized as a series of power of $r^n$ by expanding 
$\exp\{-it(V_o-V_s)\}=1 -it(V_o-V_s)+ \dots$ (since $t\simeq 1/\lQ$, $V_o-V_s \simeq mv^2$). 
Typically the nonperturbative piece 
of Eq. (\ref{vsnp}) absorbs the $\mu^\prime$ dependence of $\alpha_V$  
so that the resulting potential $V_s$ is now scale independent.\\
We notice that the leading nonperturbative term could be as important as the perturbative potential 
once the power counting is established and, if so, it should be kept exact when 
solving the Schr\"odinger  equation. In Table 1 we summarize the different kinematic situations.
\begin{table}[htb]
\makebox[2.2cm]{\phantom b}
\begin{tabular}{|c|c|l|l|}
\hline
$mv$&$mv^2$&potential&ultrasoft corrections\\\hline
$\gg \Lambda_{\rm QCD}$&$\gg \Lambda_{\rm QCD}$&perturbative& US gluons + local condensates\\
$\gg \Lambda_{\rm QCD}$&$\sim \Lambda_{\rm QCD}$&perturbative&US gluons + non-local condensates\\
$\gg \Lambda_{\rm QCD}$&$\ll \Lambda_{\rm QCD}$&perturbative + & No US (if light quarks\\
&$~$&short-range nonpert.& $\,\,\,$ are not considered)\\ 
\hline
\end{tabular}
\caption{\it Summary of the different kinematic situations.}
\label{tab2}
\vspace{-0.1cm}
\end{table}

\subsection{The perturbative potential and nonpotential 
nonperturbative corrections:\\
 Coulombic and quasi-Coulombic systems}

 Let me define what I mean with Coulombic or quasi-Coulombic systems \cite{nonlocal}.
 If
$mv\gg mv^2\simm  \lQ$, 
the system is described up to order $\alpha_{\rm s}^4$ by a potential which 
is entirely accessible to perturbative QCD (\ref{newpot}). 
Nonpotential effects start at order $\alpha_{\rm s}^5\ln \mu^\prime$ \cite{nnnll}. 
We call {\it Coulombic}
 this kind of system. Nonperturbative effects are of nonpotential type
and can be encoded into local 
(\`a la Voloshin--Leutwyler\cite{voloshin}) condensates (if $mv^2 \gg \lQ$) or non-local 
condensates (if $mv^2 \sim \lQ$):
 they are suppressed by powers of  $\lQ/mv^2$ and $\lQ/mv$ respectively. 
We will discuss in details the case of Coulombic systems in the next section.

If $mv \gg \lQ \gg m v^2$, 
 the scale $mv$ can be still integrated out perturbatively,
giving rise to the Coulomb-type  potential (\ref{newpot}).
Nonperturbative  contributions to the potential arise 
when integrating out the scale $\lQ$ \cite{pnrqcd}, precisely as explained in Sec. 6.3.
We call {\it quasi-Coulombic} the  systems  where 
the nonperturbative piece of the potential 
can be considered small with respect to the Coulombic one and treated as a perturbation.\par
Some levels of $t\bar{t}$, the lowest level of $b \bar{b}$  may be considered Coulombic systems
\footnote{Actually both
 $b\bar{b}$  and $c\bar{c}$
 ground states have been studied in this way\cite{yn}.},
while the $J/\psi$, the $\eta_c$ and the short-range hybrids may be considered quasi-Coulombic.
The   $B_c$  may be in a boundary situation.
As it is typical in an effective theory,
 only the actual calculation may confirm if the initial
assumption about the physical system was appropriate. In Sec. 9 I will give an example 
of computation in a quasi-Coulombic system after an appropriate redefinition of the mass parameter 
that subtracts the largest indetermination in the perturbative series.

For all these systems it is relevant 
to obtain a determination of the energy levels 
as accurate as possible in perturbation theory.

\section{CALCULATION of QUARKONIUM ENERGIES at $O(m\alpha_s^5 \ln \alpha_s)$}

The perturbative energy levels of quarkonium are known at $O(m \alpha_s^4)$\cite{yn}.
pNRQCD provides us with a well defined way of calculating the energy levels at higher orders,
since the size of each term in the pNRQCD Lagrangian  is
well-defined and since we are now able to calculate the static potential beyond two loops.
In order to obtain the leading logs 
at $O(m \als^5)$ in the spectrum, $V_s$ has to be computed at $O(\als^4\ln)$,
$V_s^{(1)}$ at $O(\als^3\ln)$, $V_s^{(2)}$ at $O(\als^2\ln)$ and $V_s^{(3)}$ at $O(\als\ln)$.
The matching to pNRQCD at ${\rm N}^3$LL accuracy was calculated in \cite{nnnll} and the potentials
at the requested accuracy were thus obtained. 

The total correction to the energy at $O(m\alpha_s^5 \ln\alpha_s)$ is given by the sum of the 
averaged values of the potentials plus the nonpotential contributions,
\be
\delta E_{n,l,j}= \delta^{\rm pot} E_{n,l,j}(\mu^\prime)+ \delta^{\rm US} E_{n,l}(\mu^\prime)\, ,
\label{energytotal}
\ee
\bea
&&\delta E_{n,l,j}^{\rm pot}(\mu^\prime) = E_n {\als^3 \over \pi} 
\left\{{C_A \over 3} \left[{C_A^2 \over 2} 
+4C_AC_F {1\over n(2l+1)}
+2C_F^2\left({8 \over n(2l+1)} - {1 \over n^2}\right) \right]
\ln{\mu^\prime \over m\als} \right. 
\label{energy1} \\ 
&&
\quad \quad \quad +{ C_F^2\delta_{l0} \over 3 n}\left(8\left[C_F-{C_A \over 2}\right]
\ln{\mu^\prime\over m\als} + \left[C_F+{17 C_A \over 2}\right]\ln{\als} \right)
\left. \right. \nonumber \\
& & 
\quad \quad \quad \left.  -{7 \over 3}{ C_F^2C_A \delta_{l0}\delta_{s1} \over n} \ln{\als}  
- {(1- \delta_{l0}) \delta_{s1}C_{j,l} \over l(2l+1)(l+1)n}{ C_F^2C_A \over 2} \ln{\als} \right\} ,
\nonumber
\eea
where $E_n= - mC_F^2\als^2/(4n^2)$ and $C_{j,l} = 
 -{(l+1)(4l-1)/(2l-1)}\, {\rm for}\,  j=l-1$, $=-1\, {\rm for}\, j=l$
and $={l (4l+5)/(2l+3)}\, {\rm for}\, j=l+1$.\par
The $\ln\als$ appearing in Eq. (\ref{energy1}) come from logs of the type 
$\ln{1/m r}$ in the potential. Therefore they can be deduced once the 
dependence on $\ln m$ is known. 
  The $\mu^\prime$ dependence of Eq. (\ref{energy1}) 
cancels against contributions from US energies. 
 At the next-to-leading order in the multipole expansion the contribution 
from these scales reads
\begin{eqnarray}
\delta^{\rm US} E_{n,l}(\mu^\prime) = -i{g^2 \over 3 N_c}T_F \times \int_0^\infty \!\! dt 
\langle n,l |{\bf r} e^{it(E_n-H_o)} {\bf r}| n,l \rangle \langle {\bf E}^a(t) 
\phi(t,0)^{\rm adj}_{ab} {\bf E}^b(0) \rangle(\mu^\prime), 
\label{energyNP}
\end{eqnarray}
where $H_o = {{\bf p}^2/2m} +V_o$. 

Different possibilities appear depending on the relative size of $\lQ$ with
respect to the US scale $m\als^2$. If we consider that $\lQ \sim m\als^2$ the gluonic correlator in 
Eq. (\ref{energyNP}) cannot be computed using perturbation theory. 
We are still able to obtain all the 
$m \alpha_s^5 \ln({m \alpha_s /m})$ and $m \alpha_s^5 \ln ({m \alpha_s/{\mu^\prime}})$
contributions
 where the
 ${\mu^\prime}$ dependence  cancels now against  the US contributions that have to be evaluated 
 non-perturbatively.

If we consider that $m\als^2 \gg \lQ$, Eq. (\ref{energyNP}) can be computed
perturbatively. Being $m\als^2$ the next relevant scale, the effective role of
Eq. (\ref{energyNP}) will be to replace $\mu^\prime$ by $m\als^2$ (up to finite pieces that 
we are neglecting) in Eq. (\ref{energy1}). Then Eq. (\ref{energytotal}) simplifies to 
\bea
&&\delta E_{n,l,j} = E_n {\als^3 \over \pi} \ln{\als} 
\left\{{C_A \over 3} \left[{C_A^2 \over 2} +4C_AC_F \right.
{1\over n(2l+1)}
+2C_F^2\left({8 \over n(2l+1)} 
- {1 \over n^2}\right) \right] \nonumber \\
 & & \quad 
+{ 3 C_F^2\delta_{l0} \over n}\left[C_F+{C_A \over 2}\right]
  \left.
- {7 \over 3}{ C_F^2C_A \delta_{l0}\delta_{s1} \over n} 
- {(1- \delta_{l0}) \delta_{s1} \over l(2l+1)(l+1)n}\,C_{j,l}{ C_F^2C_A \over 2} \right\} 
\label{energy2}
\eea    
plus  non-perturbative corrections that can be
parameterized by local condensates \cite{nnnll,voloshin} and 
are
of order $ \alpha_s^2 ({\lQ\over m \alpha_s})^2 ({\lQ\over m \alpha_s^2 })^{2}$ and higher.
For the $\Upsilon(1S)$ and $t\bar{t}$  non-perturbative contributions are expected not 
to exceed $100\div 150$ MeV and $10 $ MeV respectively.

The calculation  (\ref{energy2}) 
paves the way to the full ${\rm N}^3$LO order analysis of Coulombic systems and 
is relevant at least for $t\bar{t}$ production and $\Upsilon$
 physics. In the first case it is  a step
forward reaching a 50 MeV sensitivity on the top quark mass for the $t\bar{t}$ cross section 
near threshold to be measured at future Linear Colliders \cite{tt}. In the second case they improve 
our knowledge of the $b$ mass \cite{yn}.  

In particular, from (\ref{energy2}) we can estimate 
the ${\rm N}^3$LL correction to the energy level of the $\Upsilon(1S)$ and we find \cite{nnnll}
\begin{equation}
\delta E_{101} =  {1730\over 81 \pi } 
m_b \alpha_s^4(\mu) \alpha_s(\mu^\prime) \ln{1/\alpha_s(\mu^\prime)}
\simeq (80\div 100)\,  {\rm MeV},
\end{equation}
which appears not to be small.
Corrections from this ${\rm N}^3$LL terms have been calculated for the $t\bar{t}$ cross section 
and $\Upsilon(1S)$ wave function, cf. \cite{pen} and \cite{yn}. 
They also turn out to be sizeable.

Large corrections, however, already show up at NLO and
${\rm N}^2$LO and are responsible for the bad convergence of the perturbative series in terms of the pole mass.

\section{RENORMALONS, the POLE MASS and the PERTURBATIVE EXPANSION}

The bad convergence of 
the perturbative expansion 
can be, at least in part, attributed to  renormalon 
contributions.  The pole mass, thought an infrared safe quantity\cite{kronfeld},  has long 
distance contributions of order $\lQ$\cite{ren}.
Also the static potential is affected by 
 renormalons (see e.g. \cite{ren}). Rephrasing them in the 
effective field theory language of pNRQCD we can say that the Coulomb 
potential suffers from IR 
renormalons ambiguities with the following structure
\begin{equation}
V_s(r) \vert_{\rm IR\, ren} = C_0 + C_2 r^2 + \dots
\label{ren1}
\end{equation}

\begin{figure}[htb]
\makebox[0cm]{\phantom b}
\put(100,10){\epsfxsize=1.3truecm \epsfbox{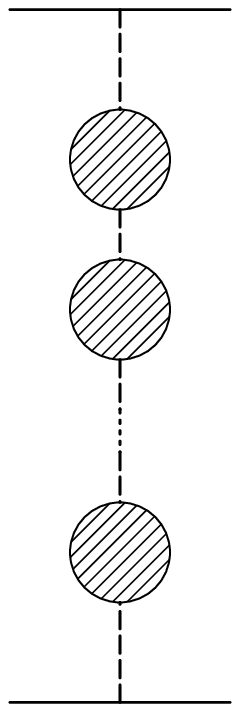}}
\put(167,55){$= C_0 =  - 2\displaystyle{ C_F  \alpha_{\rm s} (\mu) \over  \pi} 
 \mu  \sum_{n=0}^{\infty}  n! 
\left(\displaystyle{\beta_0  \alpha_{\rm s}(\mu) \over 2 \pi} \right)^n$}
\vspace{-7mm}
\caption{\it Renormalon in the static potential at LO 
in the multipole expansion, here $\mu $ is a hard cut-off, 
$1/r \gg  \mu \gg \Lambda_{\rm QCD}$.} 
\end{figure}

\begin{figure}[htb]
\makebox[0cm]{\phantom b}
\put(90,50){\epsfxsize=3.5truecm \epsfbox{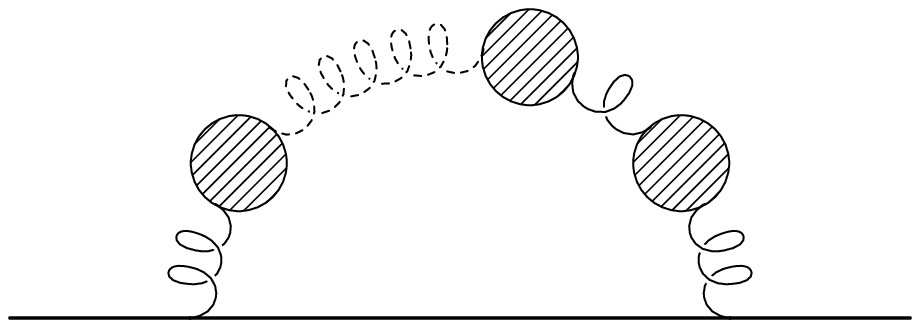}}
\put(230,55){$ =  - C_0 $}
\put(65,55){$2 \times$}
\vspace{-18mm}
\caption{\it Renormalon contribution  the pole mass. }
\end{figure}

The constant $C_0 \sim \Lambda_{\rm QCD}$ is known to be canceled by the IR pole 
mass renormalon ($2 m_{\rm pole}\vert_{\rm IR\, ren} = - C_0$, \cite{ren}). 
Several mass definitions appropriate to explicitly realize this renormalon cancellation 
 have been proposed\cite{mass}. 
Among the others, the $1S$ mass\cite{1smass} is defined as half of the  perturbative 
contribution to the $ ^3S_1$ $q\bar{q}$ mass. Unlike the pole mass, the $1S$ mass, containing,
 by construction, half of the total static energy $\langle 2m + V^{Coul}\rangle$, is free of ambiguities of order $\lQ$. Taking e.g. the $\Upsilon(1S)$, 
the $1S$ mass is related to the 
physical $\Upsilon(1S)$ mass by $E(\Upsilon(1S)) =2 m_{1S}+\Lambda_{\Upsilon} $.
$\Lambda_{\Upsilon}$  is the poorly known non-perturbative contribution, which is likely, as
we said, 
to be less than $100\div 150$ MeV.
In the next section we will present an explicit example that shows
 how,  using this mass and thus dealing with  quantities that are infrared safe at order 
$\Lambda_{\rm QCD}$, the pathologies of the perturbative series, 
due to the renormalon ambiguities affecting the pole mass, are cured.\par
It is possible to show 
 that the second infrared renormalon, $C_2 \simeq \lQ^3$, of 
$V_s$ cancels against the appropriate pNRQCD UV renormalon in   
the contribution to the potential originating at NLO in the multipole expansion. 
What remains is the  
 explicit expression for the 
operator which absorbs the $C_2 \sim \Lambda_{\rm QCD}^3$ ambiguity (for  details
 see \cite{pnrqcd}): such nonperturbative operator turns out to be a nonlocal 
chromoelectric condensate of the type used in some QCD vacuum models \cite{nonlocal,dosch}.\par
An interesting open question is if an 
explicit renormalon subtraction similar to  that one at $O(\lQ)$ can be realized
at the subsequent order $O(\lQ^3)$. This may be relevant since this renormalon is 
related to 
the corrections at ${\rm N}^3$LL discussed in the previous section. Indeed, it is
still an open problem whether the largeness of the ${\rm N}^3$LL corrections
 is an artifact due to our partial knowledge of the contributions at this order,
  or if it is an artifact  
due to the fact that the subsequent renormalon cancellation has to be realized at this order 
or finally if it  is a true signal of the breakdown of the perturbative series.
 To make more definite 
statements one  should know 
the complete ${\rm N}^3$LO or understand the mechanism of cancellation of the second renormalon \cite{sumino}.

In the next section I will present a concrete example of 
the relevance of the mass renormalon cancellation in order to obtain reliable  phenomenological 
predictions.

\section{A QUASI COULOMBIC SYSTEM with \\
EXPLICIT $O(\lQ)$ RENORMALON CANCELLATION}

I consider here  the perturbative calculation up to order $m\alpha_s^4$ of the  energy of 
the $\bar{b} c$ ground state: this will be relevant  to a QCD determination of the $B_c$ mass
 if this system is Coulombic or at least quasi-Coulombic.
 I will assume this to be the case.

In order to calculate the $B_c$ mass in perturbation theory up to order $\alpha_{\rm s}^4$, 
we  need to consider the following contributions to the potential: the perturbative 
static potential at two loops, the $1/m$ relativistic corrections at one loop,   the spin-independent 
$1/m^2$ relativistic corrections at tree level and the $1/m^3$ correction to the kinetic energy. 
We do not consider  
 $\alpha^5_s \ln{\alpha_s}$ corrections since the mechanism responsible  for this large 
contributions has not yet  been understood. 
Then, we have\cite{bc} 
\begin{eqnarray}
& &
E(B_c)_{\rm  pert} = m_b + m_c  + E_0(\bar{\mu})
\left\{\! 1 \! - {\alpha_{\rm s}(\bar{\mu})\over \pi}
 \right.  
 \left[ \beta_0  
l   + {4\over 3} C_A - {11 \over 6} \beta_0 \right] 
+ \left({\alpha_{\rm s}\over \pi}\right)^2 \left[ {3\over 4} 
\beta_0^2 \right.  
l^2   + (2 C_A \beta_0        \nonumber \\
& & 
 - {9\over 4} \beta_0^2 - {\beta_1\over 4}) 
 l 
- \pi^2 C_F^2  \left( {1\over m_b^2} + {1\over m_c^2} - {6\over m_b m_c} \right) m_{\rm red}^2
+ {5\over 4} \pi^2 C_F^2  \left( {1\over m_b^3} + {1\over m_c^3}\right) m_{\rm red}^3
 + \pi^2 C_F C_A  \nonumber \\
& & 
 + {4\over 9} C_A^2 - 
{17 \over 9} C_A \beta_0
+ \left. \left. \left( {181 \over 144} + {1\over 2} \zeta(3) +{\pi^2\over 24} \right) \beta_0^2 
+ { \beta_1 \over 4} + {c\over 8} \right]\right\} ,
\label{EBc2} 
\end{eqnarray}
being $m_{\rm red}= m_b m_c/(m_b+m_c)$,
$l=\ln({2 C_F \alpha_{\rm s} m_{\rm red}/\bar{\mu}})$,
$E_0(\bar{\mu}) = - m_{\rm red} {(C_F \alpha_{\rm s}(\bar{\mu}))^2/2}$ 
and  $\bar{\mu}$ the scale around 
which we expand $\alpha_s(r)$.

If we use here the pole masses $m_b=5$ GeV, $m_c = 1.8$ GeV and 
$\bar{\mu}=1.6$ GeV,   
then we obtain $E(B_c)_{\rm  pert} \simeq 6149$ MeV $\simeq 6800 - 115  - 183 - 353$ MeV, 
where the second, third and fourth figures are the corrections of order $\alpha_{\rm s}^2$,  
$\alpha_{\rm s}^3$ and $\alpha_{\rm s}^4$ respectively. The series turns out to be very badly
convergent.
This reflects also in a strong dependence on the normalization scale $\bar{\mu}$:
at $\bar{\mu} = 1.2$ GeV we would get $E(B_c)_{\rm  pert} \simeq 5860$ MeV, while at 
$\bar{\mu} = 2.0$ GeV we would get $E(B_c)_{\rm  pert} \simeq 6279$ MeV.
The non-convergence of the perturbative 
series (\ref{EBc2}) signals the fact that large $\beta_0$ contributions (coming 
from the 
static potential renormalon) are not summed up and canceled against the pole masses.  
In order to obtain a well-behaved 
perturbative expansion, we use, now,
the so-called $1S$ mass.   
\begin{figure}[htb]
\makebox[-1.1truecm]{\phantom b}
\put(120,0){\epsfxsize=6.6truecm\epsffile{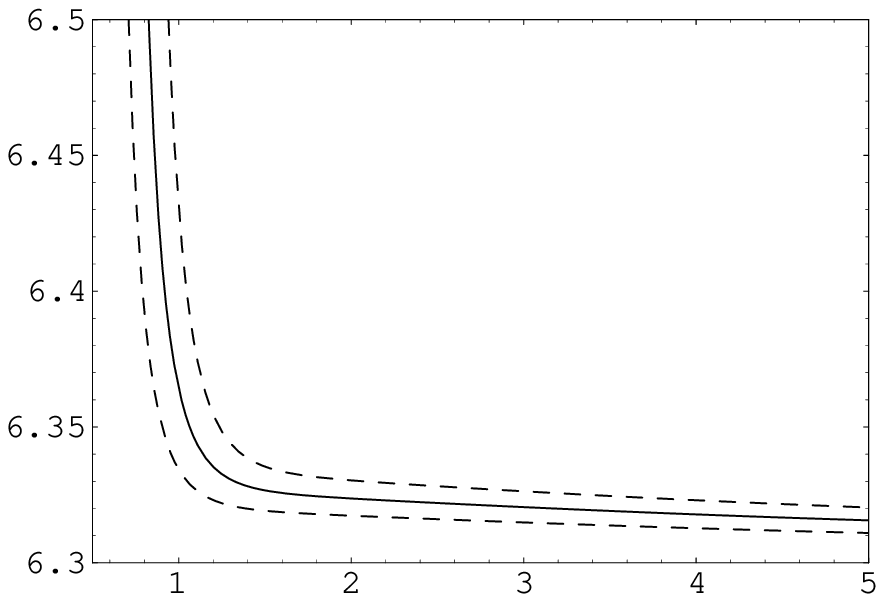}}
\put(90,50){$E(B_c)_{\rm pert}$}
\put(250,1){$\bar{\mu}$}
\caption{\it $E(B_c)_{\rm pert}$ as a function of $\bar{\mu}$ at $\Lambda_{\overline{\rm
      MS}}^{N_f=3}=300$ MeV (continuous line).  The dashed lines refer 
      to $\Lambda_{\overline{\rm MS}}^{N_f=3}=250,350$ MeV respectively.}
\label{plot}
\end{figure}
We consider the perturbative contribution (up to order $\alpha_s^4$) 
of the $\, ^3S_1$ levels of charmonium and bottomonium:
$$
E(J/\psi)_{\rm  pert}= f(m_c); \quad  E(\Upsilon(1S))_{\rm  pert}=f(m_b), 
$$
which are respectively a function of the $c$ and the $b$ pole mass and can be read off from 
Eq. (\ref{EBc2}) in the equal-mass case, adding to it the spin-spin interaction energy: 
$m(C_F\alpha_{\rm s})^4/3$. We invert these relations in order to obtain the pole masses 
as a formal perturbative expansion depending on the $1S$ masses. Finally, we insert the expressions 
$m_c=f^{-1}(E(J/\psi)_{\rm  pert})$ and $m_b=f^{-1}(E(\Upsilon(1S))_{\rm  pert})$ in  Eq. (\ref{EBc2}). 
At this point we have the perturbative mass of the $B_c$ as a function of the  
$J/\psi$ and $\Upsilon(1S)$ perturbative masses.
If we identify the perturbative masses $E(J/\psi)_{\rm  pert}$, $E(\Upsilon(1S))_{\rm  pert}$ with the physical 
ones,  
then the expansion (\ref{EBc2}) depends only on the scale $\bar{\mu}$.
 The perturbative series turns out to be reliable  
for values of $\bar{\mu}$ bigger than $(1.2 \div 1.3)$ GeV and 
lower than $(2.6 \div 2.8)$ GeV. 
For instance, $E(B_c)_{\rm  pert} = 6278.5 +  35 + 6.5 + 5.5$ MeV 
at the scale $\bar{\mu}=1.6$ GeV. 
Therefore, 
we obtain now a better convergence of the perturbative expansion and   
a stable determination of the perturbative mass of the $B_c$.
 This fact seems to support  the $B_c$ {\it being} indeed 
a Coulombic or quasi-Coulombic system.
By varying $\bar{\mu}$ from 1.2 GeV to 2.0 GeV and $\Lambda_{\overline{\rm MS}}^{N_f=3}$ 
from 250 MeV to 350 MeV and by calculating the maximum variation 
of $E(B_c)_{\rm  pert}$ in the given range of parameters, we get 
as our final result  
\begin{equation}
E(B_c)_{\rm  pert} = 6326^{+29}_{-9}\, {\rm MeV}. 
\label{ebc}
\end{equation}
As a consequence of the now obtained good behaviour of the perturbative series 
in the considered range of parameters, the result appears stable with respect 
to variations of $\bar{\mu}$ (see Fig. \ref{plot}) and,
 therefore, reliable from the perturbative point 
of view. It  represents a rather clean prediction 
of the lowest mass of the $B_c$. 
Notice that all the existing predictions are 
based either  on potential models or on lattice evaluation (with still  large errors).

Non-perturbative contributions have not been taken into account so far. 
They affect the identification of the perturbative masses 
$E(B_c)_{\rm  pert}$, $E(\Upsilon(1S))_{\rm  pert}$, 
$E(J/\psi)_{\rm  pert}$, with the corresponding physical ones through Eq.(\ref{EBc2}). 
Let us call these non-perturbative contributions $\Lambda_{B_c}$, 
$\Lambda_{\Upsilon}$ and $\Lambda_{J/\psi}$ respectively. 
As discussed before,
they can be of potential or nonpotential nature. 
In the latter
 case they can be encoded into non-local condensates or into local condensates.
  Non-perturbative contributions affect the identification with the 
physical $B_c$ mass roughly by an amount $\simeq -  {\Lambda_{J/\psi}/ 2}$ 
$- {\Lambda_{\Upsilon}/2}$  $+ \Lambda_{B_c}$. 
Assuming $|\Lambda_{J/\psi}| \le 300$ MeV, $|\Lambda_{\Upsilon}| \le 100$ MeV 
and $\Lambda_{\Upsilon} \le \Lambda_{B_c} \le \Lambda_{J/\psi}$, the 
identification of our result (\ref{ebc})
with the physical $B_c$ mass may, in principle, 
be affected by uncertainties, due to the unknown non-perturbative 
contributions, as big as $\pm 200$ MeV. However, the different non-perturbative contributions
$\Lambda$ are correlated, 
so that we expect, indeed, smaller uncertainties. If we assume, for instance, 
$\Lambda_{\Upsilon}$ and $\Lambda_{J/\psi}$ to have the same sign, which seems to be 
quite reasonable, then the above uncertainty reduces to $\pm 100$ MeV. 
 This would confirm, indeed, that 
the effect of the non-perturbative contributions on the result of 
Eq. (\ref{ebc}) is not too large.

\section{RENORMALIZATION GROUP IMPROVEMENT \\ and vNRQCD}
Of course the results presented here can be renormalization group improved,
i.e. the potentially large logarithms can be resummed. The RG group improvement 
in pNRQCD has been done only at the level of the static potential \cite{rg}
 up to now. 
When going beyond the static limit $v=0$ a special care has to be payed to the 
correct implementation of the renormalization group equations because on 
the bound state we are dealing with correlated scales $m v$ and $mv^2$ \cite{mss}.
Another EFT, called vNRQCD \cite{vnrqcd},
 has been devised explicitly to deal with the resummation of large logs.
In this case the treatment is, however, purely perturbative. 
The Lagrangian is explicitly decomposed in soft, potential 
and ultrasoft fields:
$$ 
L_{vNRQCD} = L_{s}  +  L_{p}  +  L_{u}
$$  
The  matching  is performed at the scale $ m$.
The RG equations evolve simultaneously from the scale 
$m$ to the  correlated  scales $ m v$  and $m v^2$.
This seems to provide the correct double logarithms. \par
The approach of vNRQCD may be complicated by the fact that the different scales 
remain entangled. Moreover vNRQCD is perturbative by construction and thus 
cannot be used in order to parameterize nonperturbative effects.
The goal of vNRQCD is summing up the logarithms in the velocity. 
In my opinion, however,
the real challenge in heavy quark systems 
remain the  nonperturbative corrections.
It is also  important to keep 
in mind that, as I noticed in Sec. 4, the finite part contributions may be 
as much (or more) relevant as the logarithmic   ones.

\section{RESULTS and APPLICATIONS}

In this Section we just recall  a list of physical situations where the presented 
results of  pNRQCD (in the situation of the perturbative matching $mv \gg \lQ$)
have been or may be relevant:
\begin{itemize}
\item{\it Toponium cross section \cite{tt,pen}.}
\item{\it Determination of the mass of the $b$}, once the renormalon contribution 
has been appropriately subtracted \cite{yn,pen,hoang}.
\item{\it Hybrids and gluelumps}\\ 
pNRQCD gives model independent predictions on the behaviour of the hybrid static potentials.
In particular it predicts for these  potentials an octet behaviour at very short distances 
and it correctly states all the degeneration patterns in the small $r$ region \cite{pnrqcd,hybrids}.
Moreover, it allows to relate the mass of the gluelumps to the correlation lengths of some 
nonlocal vacuum field strength correlators \cite{nonlocal,pnrqcd} allowing us to obtain 
interesting model independent information on the behaviour of these nonperturbative objects \cite{dosch}.
\item{\it Quarkonium scattering, Van Der Waals forces, Quarkonium Production.} 
Work is in progress on these applications.
\end{itemize}

\section{NONPERTURBATIVE MATCHING \\
pNRQCD for $\Lambda \simeq mv$}
In this case the potential interaction is dominated by nonperturbative effects.
This is the most interesting situation, in which most of the mesons seem to lie.\par
A large effort has been made in the last decades in order to obtain from QCD
the nonperturbative potentials in the Wilson loop approach.
pNRQCD  allows us to obtain via a nonperturbative matching all the nonperturbative 
potentials \cite{aprot}. On this respect I want to discuss few concepts and results.\par
In pNRQCD a  potential picture for heavy quarkonium 
emerges at the leading order in the US expansion 
under the condition that the matching between  NRQCD and pNRQCD  can be performed within 
an expansion in $1/m$. The gluonic excitations (hybrids and glueballs) 
that form a gap  of order $\lQ$ with respect to the quarkonium state
can be integrated out  and the potentials follow
in an unambiguous and systematic way from the nonperturbative matching to pNRQCD.
Thus, we recover the quark model from pNRQCD \cite{aprot}. The US degrees of freedom in this case are 
not coloured gluons but US gluonic excitations between heavy quarks
 and pions. They  can be systematically included 
and  may eventually affect the leading potential picture. 
Let us consider for instance the singlet matching potential.
Disregarding the US corrections, we have the identification
$$V_s(r)  = \lim_{T\to\infty}{i\over  T    } 
\ln \langle W_\Box  \rangle .$$
US corrections to this formula are due to pions and US gluonic excitations between heavy quarks,  
and may be included in the same way as the effects due to US gluons have been included 
in the perturbative situation.

The complete $1/m^2$ potential has been calculated along these lines \cite{aprot}. 
There are many appealing and interesting features of this procedure. 
The matching calculation is performed in the way of the quantum mechanical 
perturbation theory on the QCD Hamiltonian (where perturbations are counted in orders of $1/m$
and not of $\alpha_s$) and only at a later stage the relation with the Wilson loop and field insertions 
is established. This allows us to have a control on the Fock states 
of the problem and on  the contributions coming from gluonic excitations  in the 
intermediate states. At the end, all the expressions are again given in terms of Wilson 
loops, which  can be evaluated on the lattice or in QCD vacuum models.
The  potentials turn out to be naturally factorized in a hard part (the matching coefficients 
at the hard scales inherited by pNRQCD from NRQCD) and a low energy part (the Wilson loops
expressions). The power counting may turn out to be quite different from the perturbative 
(QED-like) situation and this may turn out to be very important
 for some applications 
(e.g. quarkonium production).

I will not discuss these very interesting results 
 any longer because it is the content of the review of Antonio Vairo
at this Conference \cite{aprot}.

\section{CONCLUSION and OUTLOOK}
I have shown that it is possible to construct systematically 
and within a controlled expansion an effective theory of QCD, which describes 
heavy quark bound states. All known perturbative and nonperturbative 
regimes (potential, nonpotential), are dynamically present in the theory, 
which  is equivalent  to QCD. I have presented many applications 
of pNRQCD in the situation $\lQ\ll mv$. In the situation $\lQ \simeq mv$, 
I have shortly discussed  how pNRQCD allows us to systematically factorize  the nonperturbative 
heavy quark dynamics.
Similar techniques have been applied to pionium \cite{soto}.
\vskip 0.15truecm
{\bf Acknowledgements}\par\noindent
I thank the Organizers, especially Vladimir Petrov and Sergei  Klishevich, 
for arranging such an interesting  
and nice Conference and for providing the participants with a 
particularly warm hospitality, a friendly and enjoable 
atmosphere and a perfect organization. I wish that this series of Conferences 
could continue like this  in the future. \par  
I thank Antonio Vairo for reading the manuscript and for useful comments.

\end{document}